\documentclass[prl,aps,twocolumn,superscriptaddress,showpacs,floatfix]{revtex4}
\usepackage{graphicx}
\begin{document}
\title{Gauge-invariant Geometry of Space Curves: 
Application to Boundary Curves of M\"obius-type Strips} 
\author{Radha Balakrishnan}
\affiliation{The Institute of Mathematical Sciences, Chennai  600 113,
India}
\email{radha@imsc.res.in}
\author{Indubala I. Satija}
\affiliation{Department of Physics, George Mason University, Fairfax, VA
22030}
\email{isatija@physics.gmu.edu}
\date{\today}
\begin{abstract}
We derive  gauge-invariant expressions for
the twist $Tw$ and the linking number $Lk$ of a closed space curve, that
are independent of the frame used to describe the curve, and hence
 characterize  the  intrinsic geometry of the  curve. We are thus led to  
 a {\it frame-independent} version  of the  
C\u{a}lug\u{a}reanu-White-Fuller theorem $Lk =Tw + Wr$ for a curve, 
where $Wr$ is the writhe of the curve. 
The gauge-invariant twist and writhe are related to two types of 
geometric phases associated with the curve. 
As an application, we
  study the geometry of 
the boundary curves of closed twisted strips.
  Interestingly, the M\"obius strip geometry 
is singled out by a characteristic maximum that 
appears in the geometric phases, at a 
certain critical width of the strip.  
\end{abstract}
\maketitle

Construct a M\"obius strip (MS) by taking a  long rectangular strip
of paper, 
making a half-twist at one of its ends, and bringing the ends together. 
This strip has only one edge or 
boundary curve, which is a closed twisted space  curve.     
Now, keeping the width constant,  
slide one of the ends on top of the other,  
so as to effectively decrease 
the length of the strip, and watch how the geometry of the 
boundary curve changes. 
Once the ratio of the width of the strip to its length 
increases to a certain value, a  sharp twist appears  
on the boundary, which soon develops into a
cone-like singularity.   
Since the boundary curve of the strip encodes information 
about this ratio,  a key question is
how the geometry of the curve changes with this parameter.
Answering this is important in view of the fact that  diverse systems, 
such as biopolymers\cite{fran}, 
vortex filaments in a fluid\cite{hasi}, magnetic field lines\cite{moff},
twisted optical fibers\cite{chia}, 
and phase space trajectories in dynamical systems\cite{bala1}, 
can be modeled using
closed twisted space curves. To this end, 
we first present a novel framework to describe the 
geometrical characteristics of 
 a general twisted space curve, and then 
apply it to the  boundary curve
of the MS, as well as to other strips  related 
to the MS, i. e., strips with several 
half-twists.

A basic concept  in the study of the geometry of a space curve  is 
that of a thin ribbon\cite{full}, one of whose edges is
the given curve. To construct the other edge of the ribbon, 
one must first  choose 
an orthogonal frame  to describe the curve. 
 A result  which is widely applied  
is the  celebrated  Cal\u{u}gareanu-White- Fuller (CWF) theorem 
for a  closed  ribbon\cite{full, calu, whit,pohl}. 
It states that the integer linking number $Lk$ 
of the two edges  of the ribbon can be written as the sum of the twist  
 $Tw$ of the  ribbon  and the   writhe $Wr$ of the  space curve.   
 $Tw$ characterizes 
 the   twist of the ribbon
about its axis, while  $Wr$ is a measure of the non-planarity and 
non-sphericity\cite{denn} of the space curve.

Now, given a space curve, an infinite number of 
ribbons can be associated with it,
owing to the inherent freedom in the choice of the orthogonal frame used.
Each frame will yield a certain value for $Tw$, and hence for $Lk$.
It is therefore desirable to find a way to define
an {\it intrinsic} twist $Tw_g$ of a curve about its axis, 
which would have a unique value,
independent of the frame used.
In this Letter, we achieve this by invoking the principle 
of gauge invariance. 
This leads us to the definition of 
an {\it intrinsic} linking number $Lk_g$ of a space 
curve. In short, we  obtain  a ribbon-independent, 
gauge-invariant version of the CWF theorem 
for a  closed  space curve, given by
\begin{equation}
Lk_g=Tw_g+Wr.
\label{CWFG}
\end{equation} 
We also show  that $Tw_g$
can be viewed as a geometric phase or an anholonomy 
associated with the space curve. 
This classical geometric phase  bears a close analogy to the  quantum
geometric phase obtained in the Aharonov-Anandan 
formulation\cite{ahar}. 
Further, we show that $Wr$  can also be interpreted as 
another distinct type of geometric phase, that can be associated with a 
certain {\it moving} space curve\cite{bala2}. It 
follows that the linking number 
$Lk_g$ of a closed space curve can be viewed as 
the difference between two distinct
geometric phases associated with the curve.

Our analysis has a distinct advantage over the usual 
ribbon formulation\cite{full} in the 
geometric characterization of curves
with inflexion points, i. e.,
points where the curvature vanishes. 
Although curves with inflexion points are common in many applications,
one assumes their absence in the standard formulation. 
In contrast, while  $Tw$ and $Lk$  change discontinuously   
by unity whenever the curve develops an inflexion point,
 $Tw_g$ and $Lk_g$  remain unaffected. Hence the relationship (\ref{CWFG})  
  holds even when the curve has inflexion points, which is 
   a very desirable feature in applications.
To illustrate this, we present a systematic study of 
the change in geometry of the boundary curve of 
the M\"obius strip, as well as those of other
 multi-twisted  strips, as the strip-width varies.   These curves
 typically acquire  inflexion points at a certain critical strip-width.

We begin by developing a frame-independent formalism for describing the 
geometrical characteristics
of a {\it general} closed space curve $C_{1}={\bf r}(s)$ of length $L$, 
parameterized by the arc length $s$.
The unit tangent vector to the curve is ${\bf T} (s) ={\bf r}_{s}$ where
the subscript $s$ stands for the derivative with respect to $s$.
Let {\bf U} be a unit vector perpendicular to the curve, so that 
${\bf T}, {\bf U}$ 
and ${\bf V}={\bf T}\times{\bf U}$  define an orthogonal frame
at every point on the curve. 
Since ${\bf U}$ must lie on the plane perpendicular to the tangent,
there is an inherent freedom  in choosing 
its direction on this plane, signifying a certain
{\it gauge freedom} in its choice.
 
As mentioned in the beginning, the ribbon is bounded by 
$C_1={\bf r}(s)$ and a neighboring curve 
$C_2={\bf r}(s)+\epsilon {\bf U}(s)$, where $\epsilon$ is small.
The  twist $Tw$ of the {\it ribbon} is defined\cite{full} as 
$Tw({\bf r},{\bf U})=
=(2\pi)^{-1}\int_{0}^{L}
{\bf V} \cdot {\bf U}_{s}\,ds.$ 
Hence $Tw$ depends on the direction of ${\bf U}$ chosen, 
and is a frame-dependent
quantity. Defining the complex unit vector 
${\bf Q}=({\bf U}+i{\bf V})/\sqrt{2}$, we get
\begin{equation} 
Tw({\bf r},{\bf U}) 
=\frac{1}{2\pi}\int_{0}^{L}
({\bf T}\times{\bf U}) \cdot {\bf U}_{s} ds=
\frac{i}{2\pi}\int_{0}^{L} {\bf Q^*}\cdot {\bf Q}_s ds.
\label{Tw}
\end{equation}
Since we seek frame-independent geometrical 
characteristics of the curve, 
we make a gauge transformation ${\bf Q} \rightarrow {\bf q}=
({\bf u}+i{\bf v})/\sqrt 2={\bf Q} e^{i\eta(s)}$, 
which is essentially the rotation
of the frame about the tangent, 
by an angle $\eta(s)$. It is easy to verify that 
${\bf q}^*\cdot{\bf q}_s= {\bf Q}^*\cdot{\bf Q}_s~+~i~\eta_{s}$.
On integration, this leads to
\begin{equation}
i\int_{0}^{L}{\bf q}^*\cdot{\bf q}_{s}~~ds =
i\int_{0}^{L}
{\bf Q}^*\cdot{\bf Q}_{s}~~ds ~ +~[\eta(0)-\eta(L)].
\label{integ}
\end{equation}
Since 
$[\eta(0)-\eta(L)] = {\rm arg}\,[{\bf q}^*(0)\cdot{\bf q}(L)]-
{\rm arg}\,[{\bf Q}^*(0)\cdot{\bf Q}(L)]$, we see that
\begin{eqnarray}
Tw_g({\bf r})
&=&\frac{i}{2\pi} \int_{0}^{L}{\bf Q}^*\cdot{\bf Q}_{s}~~ds
~-~~\frac{1}{2\pi} {\rm arg}
\,[{\bf Q}^*(0)\cdot{\bf Q}(L)]\nonumber\\
&=&Tw({\bf r}, {\bf U})-
\phi_{T}({\bf r},{\bf U})
\label{Twg}
\end{eqnarray}
 is independent of ${\bf U}$, and may 
therefore be identified with 
the {\it gauge-invariant twist}
of the curve.
In Eq. (\ref{Twg}), the second term $\phi_{T}({\bf r},{\bf U})$
is just the total phase (in units of $2\pi$)
accumulated by  ${\bf Q}$ as the frame 
moves from $0$ to $L$ on the curve, and is given by   
\begin{equation}
\phi_{T}({\bf r},{\bf U})
=\frac{1}{4\pi}{\rm tan}^{-1}\left(
\frac{{\bf U}(L)\cdot{\bf V}(0)-{\bf V}(L)\cdot{\bf
U}(0)}{{\bf U}(L)\cdot{\bf U}(0)+{\bf V}(L)\cdot{\bf V}(0)}\right).
\label{arctan}
\end{equation}
Up to this point, 
the formulation is valid for both open and closed space curves.
For the latter, the total phase $\phi_T$ is always an integer.

The twist $Tw$ also appears in the following CWF theorem for a ribbon
associated with a closed space curve  ${\bf r}(s)$:
\begin{equation}
Lk({\bf r}, {\bf U})=Tw({\bf r}, {\bf U})+ Wr({\bf r}),
\label{CWF}
\end{equation}
 where the linking number $Lk({\bf r}, {\bf U})$ 
of the curves $C_1$ and $C_2$ 
depends on the choice of ${\bf U}$, 
and $Wr$ is the writhe of the curve.
Following Dennis and Hannay\cite{denn}, 
we find it convenient to cast  
the usual expression\cite{full} for $Wr$
in terms of the chord ${\bf C}$, as follows:
\begin{equation}
Wr=\frac{1}{4\pi} \int \!ds_1 \int \!ds_2 
\,\left(\frac{\partial{\bf C}}
{\partial{s_1}}\times\frac{\partial{\bf C}}
{\partial{s_2}}\right)\cdot{\bf C}.
\label{WR}
\end{equation}
Here  ${\bf C}(s_1,s_2)=[{\bf r}(s_1)-{\bf r}(s_2)] 
/ |{\bf r}(s_1)-{\bf r}(s_2)|$ 
is a unit vector, and 
$s_1\,,\,s_2$ are two points on the 
given curve ${\bf r}(s)$. Thus $Wr$ 
is a frame-independent, intrinsic property of the 
space curve, and is gauge-invariant.

 Combining Eqs. (\ref{Twg}) and (\ref{CWF}),  we obtain
$Lk({\bf r},{\bf U})-\phi_{T}({\bf r},{\bf U})=Tw_g+Wr$. 
The gauge-invariance of $Tw_g$ and $Wr$ implies that 
\begin{equation}
Lk_{g}({\bf r})=Lk({\bf r},{\bf U})-\phi_{T}({\bf r},{\bf U})
\label{Lkg}
\end{equation}
 is independent of ${\bf U}$. $Lk_g$ will be called  
the {\it gauge-invariant linking number} of the curve.
 Combining  Eqs. (\ref{Twg}) and (\ref{Lkg}), 
we obtain Eq. (\ref{CWFG}), the 
 gauge-invariant form of the CWF theorem. 

The advantage of the gauge-invariant formulation is  
that the expressions for $Tw_g$ (and $Lk_g$) can now be 
written down in {\it any} frame. It is convenient to  
 use  the Frenet-Serret (FS) frame\cite{stru}, 
where ${\bf U}$ and ${\bf V}$ are the principal 
normal ${\bf N}={\bf T}_s/\vert {\bf  T}_s\vert$ and 
the binormal ${\bf B}={\bf T}\times{\bf N}$, 
respectively. The 
frame rotation is given by the FS equations, 
${\bf T}_s=\kappa {\bf N}, {\bf N}_s=-\kappa {\bf T}+\tau {\bf B}$
and ${\bf B}_s=-\tau {\bf N}$,  
 where $\kappa=\vert {\bf T}_s\vert$ is the curvature  
and $\tau~
=~{\bf T}\cdot({\bf T}_s\times{\bf T}_{ss})/
({\bf T}_s\cdot{\bf T}_{s})$ is the torsion.
Setting ${\bf U}={\bf N}$ in   Eqs. (\ref{Tw}) and (\ref{arctan}), 
and substituting into Eq. (\ref{Twg}), 
we get 
\begin{equation}
Tw_g = \frac{1}{2\pi}\int_{0}^{L} \tau \,ds-\phi_{T}({\bf r},{\bf N}).
\label{TwgFS}
\end{equation}
In what follows, we show that  $Tw_g$ can be viewed as a 
geometric phase or an  anholonomy 
associated with the space curve. Using the FS equations, 
we see that the first term 
in Eq. (\ref{TwgFS}), i. e.,
the integrated torsion, is the 
rotation of the $({\bf N},{\bf B})$ plane
about the tangent, as measured with respect to a local inertial
 (non-rotating) frame, as it moves once around  the
curve. This is called Fermi-Walker transport\cite{kugl}.  
This term can be regarded as the {\it dynamical} 
phase accumulated by the 
$({\bf N},{\bf B})$ plane.
The second term  in Eq. (\ref{TwgFS}) is obviously the corresponding 
{\it total} phase.  
Therefore, the gauge-invariant 
geometric phase $\phi_g$ associated 
with the $({\bf  N},{\bf B})$ plane rotation
 on the curve is just the difference of the 
two phases above, i. e., $\phi_g=\phi_{T}({\bf r},{\bf N}) -  
 (2\pi)^{-1}\int_{0}^{L} \tau \,ds$. 
On using Eq. (\ref{TwgFS}), 
we get $\phi_g=-Tw_g$,
showing that the gauge-invariant twist $Tw_g$ is 
just the negative of the geometric 
phase $\phi_g$ (in units of $2\pi$). This result is valid for both open and closed curves. 
We refer to this 
as tangent anholonomy, since the plane involved
is perpendicular to the tangent.

Next, we show that the writhe $Wr$ is related 
to another type of  
geometric phase. In  
Eq. (\ref{WR}) for $Wr$, the unit
chord ${\bf C}$ is a function of
two independent variables, $s_1$ and $s_2$. Thus ${\bf C}$ 
can be regarded as a  unit tangent vector to
a  certain space curve parameterized by 
$s_1$ for a given $s_2$, and which  moves as $s_{2}$
varies. Therefore we can use the frame evolution formalism
for a moving space curve\cite{bala2}. Using the idea of Fermi-Walker
transport once again,
it can be shown that $Wr$ is geometric phase 
associated with  the rotation of the plane perpendicular
to the chord ${\bf C}$.
Hence  $Wr$ can be viewed as a chord anholonomy.
These results show that $Lk_g$, which is an integer, 
is the difference
between the chord anholonomy and the 
tangent anholonomy, which are not necessarily integers.

We now apply the gauge-invariant 
CWF theorem to study the geometry
of the boundary curves of 
twisted closed strips. 
These strips are surfaces described by
\begin{eqnarray}
x&=&\left(R+w \cos\frac{nt}{2}\right)\cos t,\,\,
y=\left(R+w \cos\frac{nt}{2}\right)\sin t, \nonumber \\ 
z&=&w\sin\frac{nt}{2},
\label{ms}
\end{eqnarray}
with parameters $-\alpha\le w \le \alpha $ and $0\le t \le 2\pi$.
The length of the strip is $2\pi R$, and we set $R=1$ for convenience.
 $\alpha$ is the half-width of the strip, and
the integer $n$ is the number of half-twists on the strip (so that
$n=1$ refers to the well-known M\"obius strip). 
The boundary of the strip  
corresponds to $w=\pm \alpha$ in Eq. (\ref{ms}).
For odd $n$, the strip is a  non-orientable surface with one boundary
curve, so that the range of $t$ is $[0,4\pi]$. 
For even $n$, the strip is orientable, 
with two boundary curves
with similar geometries, and 
the range of $t$ is $[0,2\pi]$ for each curve.
 Topologically, for all widths, the 
boundary curves of  the odd-$n$ strips  with $n > 1$
are knotted curves 
(e. g, it is a trefoil knot for $n=3$, a 
five-pointed star knot for $n=5$, etc.), 
while those of the even-$n$ 
strips are not knotted. The boundary curve of the 
MS clearly does not fall 
in either of these classes, since it is the only
case in which the  boundary curve
 of a non-orientable strip is not a knot. 
For all $n$, the geometry of the boundary curve 
of the strip has a non-trivial dependence
on its width, as we shall show explicitly. 

\begin{figure}[htbp]
\includegraphics[width=2.7in, angle=270]{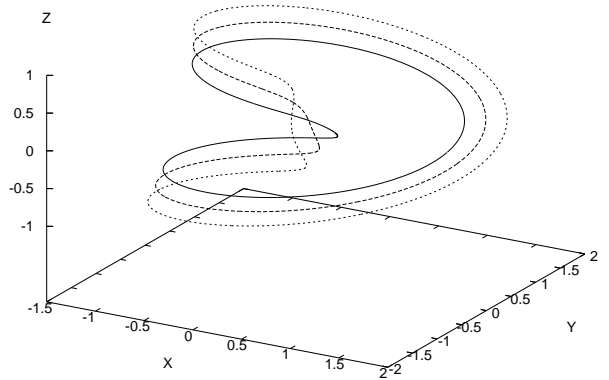}
\leavevmode
\caption{Boundary curves of  M\"obius strips with 
half-widths  $\alpha=0.6, 0.8$ and $1.0$.
The middle curve is the critical curve with $\alpha=0.8$, 
and the local straightening of this
curve near the inflexion point is clearly visible.}
\label{bound}
\end{figure}

Symbolic manipulation\cite{wolf} facilitates 
the analytic calculation of curvature and torsion
of the boundary curves of the strips given by 
Eq. (\ref{ms}). For each $£n$, 
when the half-width attains a 
critical value
$\alpha_c = =(1+\textstyle{\frac{1}{4}}n^2)^{-1}$,  
the curvature vanishes at $n$ points on the 
boundary curve\cite{rade}. For the MS,
the critical boundary curve, corresponding to 
a half-width $\alpha_c= \textstyle{\frac{4}{5}}$, 
has a single inflexion point (labeled by $t = t_c$). 
Plotting the  boundary curves of the  MS with various 
half-widths
in Fig. \ref{bound},
we see  that the  critical curve at the above $\alpha_c$ is
 distinguished by a local `straightening' of the 
curve around the point $t_c=2\pi$.

A Taylor expansion of $\kappa(t)$ and $\tau(t)$  
near $\alpha_c$ and $t_c$ gives
$\kappa \simeq 
[(\alpha-\alpha_c)^2+9b^2v_c^2(t-t_c)^2]^{\frac{1}{2}}
v_c^{3/2}$
and $\tau \simeq -3b(\alpha-\alpha_c)/
[(\alpha-\alpha_c)^2+9b^2v_c^2(t-t_c)^2]$,
where  $b=\frac{6}{5}$ and $v_c = | d{\bf r}/dt |_c = 1/\sqrt{5}$. 
These expressions show that at $t=t_c$, $\tau$ is singular 
at $\alpha_{c}$, and changes sign as the parameter $\alpha$
passes through $\alpha_c$.
We point out that
the functional form of $\kappa$ and $\tau$ 
near the inflexion point is identical to 
that obtained in an earlier study\cite{moff}, which
investigated the behavior of a curve in the 
vicinity of an inflexion point,
the only difference being the value of $b$, 
which was unity for the illustrative example 
of the curve chosen there.

To find the effect of the above singularity in $\tau$, 
  we calculate the angle of twist of the 
$({\bf N},{\bf B})$ plane over an interval 
$2t_0$ centered about $t_c$: this is
$\int_{-t_{0}+t_{c}}^{t_{0}+t_{c}} \tau\, v\, dt = 
\tan^{-1}\big(t_0/[bv_c(\alpha-\alpha_c)]\big)$.
As we pass through $\alpha_c$, this angle 
changes from $-\pi$ to $\pi$ 
(irrespective of the value of $t_0$), 
giving rise to a jump of $2\pi$, 
as discussed in Ref. \cite{moff}.   

Interestingly, the  above jump in the integrated 
torsion resulting from 
an inflexion point is accompanied by an identical 
jump in $\phi_{T}(\bf{r}, \bf{N})$. 
A Taylor expansion of ${\bf N}$  near
 inflexion point yields 
 ${\bf N}\simeq \kappa^{-1}
[a_{1}(t-t_c),a_{2}(\alpha-\alpha_c),
a_{3}(\alpha-\alpha_c)]$,
 where $a_{i} \,(i=1,2,3)$ are constants. 
 This  shows that at $t=t_c$  for $\alpha\rightarrow \alpha_c- 0$, 
${\bf  N}$ rotates by $\pi$
 about the tangent, while 
 for $\alpha \rightarrow \alpha_c+ 0$, it rotates by $-\pi$.
The same is also true for ${\bf B}$. Therefore, 
the angle of rotation of the
 $({\bf N}, {\bf B})$ plane increases by $2\pi$ 
as we pass the inflexion point.
As a result, $Tw_g$ computed from Eq. 
(\ref{TwgFS}) does not show a jump 
if the curve develops an inflexion point. 
The same behavior holds good for $Lk_g$.

\begin{figure}[htbp]
\includegraphics[width=2.5in]{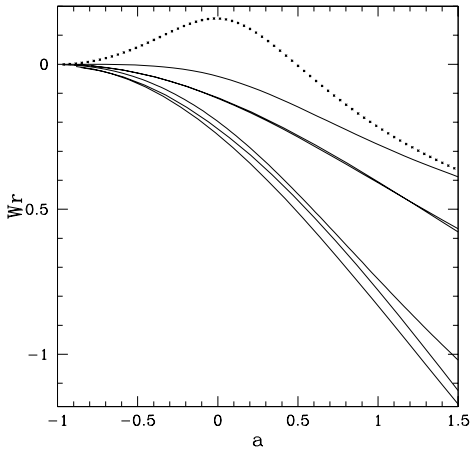}
\leavevmode
\caption{ Dependence of $Wr$ of boundary curves of 
various $n$-twisted strips (shifted down by $n$
for all odd $n$)  
on  $a=(\alpha-\alpha_c)/\alpha_c$. The top curve 
corresponds to the M\"obius strip with $n=1$,
followed by three curves that correspond to  $n=2,4,6 $ 
(top to bottom). 
The three other curves below these correspond to  $n=3,5,7$.}
\label{writhe}
\end{figure}

Next, we report the results for the 
dependence of $Wr$ and $Tw_g$  
of the boundary curves 
of the  closed twisted strips (Eq. (\ref{ms})) 
on the strip width. 
  $Wr$ of the boundary curve was directly computed from Eq. (\ref{WR}) using numerical
 integration. 
In Fig. \ref{writhe}, we plot the dependence of the writhe 
$Wr$ of the boundary curves 
of various twisted strips on the dimensionless parameter 
$a=(\alpha -\alpha_c)/\alpha_c$.  We find that for 
the boundary curves of infinitesimally thin narrow-width 
strips with $\alpha\simeq 0$,
$Wr=n$ for all odd $n$, while $Wr=0$ for all even $n$.  
We have therefore plotted, for convenience, a shifted writhe
$(Wr-n)$ for $n=1,3,5,7 $.
For $n=2, 4, 6 $, $Wr$ has been plotted without such a shift.

As seen in Fig. \ref{writhe},  $Wr$ does not undergo jumps, 
and remains smooth as the 
boundary curve develops  inflexion points at a 
corresponding critical width. 
The  variation of  $Wr$ 
with strip-width  
is significantly larger for the odd $n$, non-orientable class of strips, 
as compared to the even-$n$, orientable class. 
Another notable feature  is that
the MS boundary curve gets singled out due to its 
characteristic {\it maximum}  near the critical 
 width. This is a reflection of the fact that  the MS boundary curve 
with $n=1$ is indeed special, as it belongs to a 
class distinct from the rest, as we have discussed above.

Analytical expressions for $\tau$, ${\bf N}$ and ${\bf B}$ were used
to calculate $Tw_g$ from Eq. (\ref{TwgFS}).
The variation of $Tw_g$ with $a$  for various $n$ values 
turns out to be essentially complementary to the variation
of  $Wr$ given in Fig. \ref{writhe}, and hence is not given here. 
It is sufficient to note that
$Tw_g$  for the MS boundary curve has a characteristic {\it minimum}, 
and so the geometric phase $\phi_g=-Tw_g$ has a 
{\it maximum}. By adding $Wr$ and $Tw_g$, we find that  
$Lk_g= (n+2)$ for odd values of $n$, while  
$Lk_g=1$ for  even values of $n$.

The general CFW theorem Eq. (\ref{CWFG})  
relates topological and geometric properties of a space curve, and 
would  therefore find many applications in physics and biology. 
For instance, the two boundary curves of an orientable twisted strip 
can represent  the
two strands of a closed DNA\cite{fran}, the distance between them being 
the strip width. 
Our results may also find application
in the fabrication of biomaterials using molecular
architecture\cite{zhan}, where information about
the conformational changes in the strands caused by the variation 
of width-to-length  ratios, number
of twists, etc., would be useful. 

Since the geometric phase  describes 
the change in the polarization 
of light\cite{chia} as it propagates along 
a nonplanar circuit,
our results
 on the dependence of the geometric phase $\phi_g$ 
 on the width-to-length ratio
of the strip  can be exploited 
to tune polarizations.  
 This can be achieved by fabricating 
 an optical fiber glued to  the boundary 
of an opaque twisted strip, and studying light propagation through 
the fiber.

RB is an Emeritus Scientist, CSIR (India).

\end{document}